\documentclass[%
 reprint,
nofootinbib,
aps,
]{revtex4-2}
\pdfoutput=1
\usepackage{graphicx}% Include figure files
\usepackage{dcolumn}
\usepackage{bbold}
\usepackage{comment}
\usepackage{physics}
\usepackage{multirow}
\usepackage{diagbox}
\usepackage[utf8]{inputenc}
\usepackage[english]{babel}
\usepackage[T1]{fontenc}
\usepackage{amssymb,amsmath,amsthm}
\usepackage{hyperref}
\usepackage{physics}
\usepackage{multirow}
\usepackage{colortbl}   
\usepackage{diagbox}
\usepackage{bm}% bold math
\usepackage{hyperref}% add hypertext capabilities
\usepackage{tikz}
\usepackage{lipsum}
\usepackage{soul}

\newcommand{\user}{\textsf{User}}
\newcommand{\users}{\textsf{Users}}
\newcommand{\autho}{\textsf{Authorizer}}
\newcommand{\distrib}{\textsf{Distributor}}

\newcommand{\niter}{N}

\newtheorem{theorem}{Theorem}
\newtheorem{lemma}[theorem]{Lemma}

\begin{document}

\title{Device-independent quantum authorization based on the Clauser-Horne-Shimony-Holt game}

\author{Ricardo Faleiro}
\email{ricardofaleiro@tecnico.ulisboa.pt;}
\author{Manuel Goul\~ao}
\email{mgoulao@math.tecnico.ulisboa.pt}
%\thanks{\\(The authors are presented in alphabetical order.)}
\affiliation{Instituto de Telecomunicaç\~oes and Departamento de Matem\'atica  Instituto Superior T\'ecnico, Avenida Rovisco Pais 1049-001, Lisboa, Portugal}

\begin{abstract}
In the spirit of device-independent cryptography, we present a two-party quantum authorization primitive with non-locality as its fueling resource.
Therein, users are attributed authorization levels granting them access to a private database accordingly. The authorization levels are encoded in the non-local resources distributed to the users, and subsequently confirmed by their ability to win CHSH games using such resources. We formalize the protocol, prove its security, and frame it in the device-independent setting employing the notion of CHSH self-testing via simulation. 
Finally, we provide a proof-of-concept implementation using the Qiskit open-source framework.
   
\end{abstract}
\maketitle
\section{Introduction}
Non-locality is arguably one of the most distinct features of quantum mechanics. In a nutshell, by exploiting the fact that quantum states can be entangled, spatially separated agents sharing such entangled states can generate classically irreproducible correlations, thus breaking a cornerstone of classical reasoning, i.e. local realism \cite{B64}. Non-locality has proved to be not only  fundamental feature of nature but also useful one --- enabling a variety of different applications, in quantum computation, quantum communication, and of course \textit{quantum cryptography}. 
In the latter example, whether considering plain quantum key distribution (QKD) (e.g. E91 \cite{E91}) or more elaborate approaches, such as secure delegated quantum computation \cite{B16}, non-locality seems a prevalent motif within quantum cryptographic applications. In fact, non-locality has proved to be an essential tool regarding the notion of \textit{self-testing}, first introduced by Mayers and Yao \cite{MY04, MY98}, wherein quantum systems can be unequivocally certified only by means of observing classical statistics offered by (even) untrusted measurement devices. 
In general, self-testing is useful in order to make quantitative claims about certain features of quantum systems, for instance, \textit{``how much entanglement does a particular quantum state have?''}.
Self-testing is also intimately related to device-independent cryptography ---  a branch of quantum cryptography where security is based on properties one can infer by the classical statistics generally drawn from such self-tests. Device-independent cryptography is heralded as a powerful approach in proving security for quantum cryptographic primitives. Some  examples found in the literature are: QKD \cite{VV14, PABGMS09}, bit commitment \cite{AMPS16, SC11}, random number generation \cite{P10, AMPS12}, position verification \cite{RT18}, and coin flipping \cite{SC11,AC11}.

It is precisely following that spirit that we introduce a quantum authorization protocol which although conceptually simple, exhibits the fundamental capabilities that non-locality allows in the construction of secure quantum protocols.
This primitive is intimately related to the notion of private access to a database, which is a common task of relevant study in the field of two-party cryptography \cite{KW16}.

\section{Non-local Games}
\begin{center}
\begin{figure}%[h!]
\includegraphics[scale=.5]{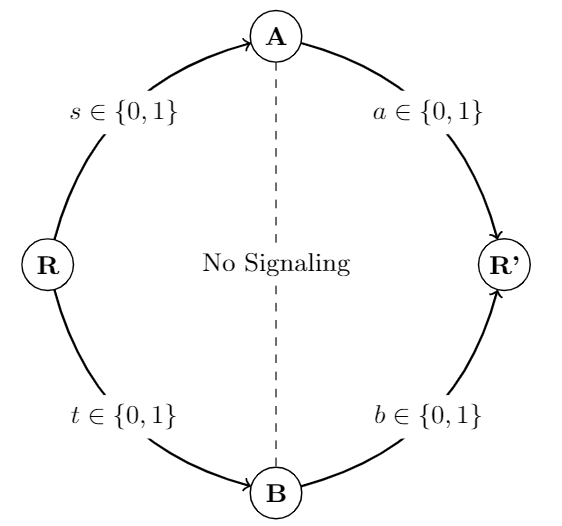}
  \caption{Interactive picture of the CHSH game being played between Alice (A) and Bob (B) mediated by the Referee; $s$ and $t$ are the questions asked to Alice and Bob, respectively, and $a$ and $b$ are their answers to the Referee at a later time (R'), when it evaluates if the winning condition $s\cdot t=a\oplus b$ has been met.}
  \label{NLG}
\end{figure}
\end{center}

Non-local games offer a powerful framework to study non-locality by analysing the probabilities of success that distributed players have when playing games where communication is explicitly prohibited --- this setup is also sometimes denoted as, \textit{Bell scenario}. We are going to use this framework in the present work, and as such we give a brief conceptual introduction to it, with a particular focus on the Clauser-Horne-Shimony-Holt (CHSH) \cite{CHSH69} game which will be explicitly used in the quantum authorization protocol --- nevertheless we refer the reader to more comprehensive technical introductions \cite{B14,CHTW04}. A non-local game $G$ is characterized by the set of possible \textit{inputs} the players can receive, the set of \textit{outputs} they can reply, a probability distribution over the set of \textit{inputs} ($p$), and some winning condition ($W$) relating the inputs and outputs. The CHSH game is a two player game where the players (Alice and Bob) try to output \textit{bits} $a$ and $b$ such that their outputs respect the winning condition, $s\cdot t=a\oplus b$, for uniformly distributed input bits $s$ and $t$, communicated individually to them by a neutral party, the Referee (see Figure \ref{NLG}).
It is known that the best classical strategy  Alice and Bob can implement only achieves a winning probability of ${3}/{4}$, thus setting the classical Bell-bound. That bound is then broken by allowing entanglement to be shared between the players, which sets the quantum Tsirelson-bound \cite{T80} of $\cos^2{({\pi}/{8})}$.  It was shown that the best quantum strategy for the CHSH is implementable with one \textit{ebit} of information and local projective measurements \cite{CHTW04} --- in practice, this means that the players need to share an Bell pair, i.e.\ a pair of qubits in any of the maximally entangled states (Bell states), 
\begin{equation}
    \ket{\phi^{\pm}}=\frac{1}{\sqrt{2}} \big(\ket{00}\pm\ket{11}\big);\;
    \ket{\psi^{\pm}}=\frac{1}{\sqrt{2}} \big(\ket{01}\pm\ket{10}\big).
\end{equation}

Nonetheless, non-maximal  violations of the Bell bound are also a possibility, for instance, if Alice and Bob are playing the game with non-optimal resources having less than 1 \textit{ebit} of information. In \cite{V02}, the authors found that there is a straightforward relationship between the maximum violation of the CHSH game/inequality and the amount of entanglement shared in the used resource, measured by the \textit{concurrence} ($\mathcal{C}$) of the state.
Thus, in such a case, one can write the optimal probability of winning the CHSH game ($\omega$) explicitly as a function of $\mathcal{C}$, as
\begin{equation}
    \label{prob}
    \omega(\mathcal{C}) = \frac{1}{2}+\frac{1}{4} \sqrt{1 + \mathcal{C}^2}, 
\end{equation}
where for some state $\rho$ the concurrence is defined as $\mathcal{C}\equiv \textup{max}\{0,\sqrt{\lambda_1} - \sqrt{\lambda_2} -\sqrt{\lambda_3}-\sqrt{\lambda_4}\}$, with $\lambda_i$ being the decreasingly
ordered  eigenvalues of $\rho (\sigma^y\otimes\sigma^y) \rho^* (\sigma^y\otimes\sigma^y)$, and $\sigma^y$ the Pauli $y$-matrix.
For some $\rho$, $ 0 \leq \mathcal{C}(\rho)\leq 1$, where, if $ \mathcal{C}(\rho)=0$, the state is separable, and $ \mathcal{C}(\rho)=1$ maximally entangled.

In a non-local Bell scenario, it is standard to assume that Alice and Bob prepare their own resources before playing the game, as that also comprises part of their strategy. As such, one reasonably accepts they will always prepare a Bell state to play the CHSH game, since they know it to be the optimal quantum resource.
On the other hand, if the resources cannot be prepared by the players themselves, they could be distributed to them (by the Referee) before the game starts.
Let us now consider the possibility that the Referee does not necessarily distribute optimal quantum resources but, say, arbitrarily entangled two-qubit pure states, parameterized by $\theta$ in their Schmidt form,
\begin{equation}
    \label{nonlocal}
    \ket{\psi_{\theta}}=\cos{\theta}\ket{00}+\sin{\theta}\ket{11},\; \textup{for}\; \theta \in [0,\pi/4].
\end{equation}
Then, Alice and Bob are  forced to use such distributed resources
since, unfortunately for them, they are restricted by the \textit{no signalling condition} which is at odds with the LOCC (local operations and classical communication) paradigm.
This precludes the possibility of \textit{Entanglement Distillation} \cite{BBPS96} by restricting them to the  LO (local operations) class --- and since local operations cannot  deterministically increase the entanglement they already have \cite{NC02}, in this scenario the distributed resource is the best one they have to play the game, even if not the optimal one.

Considering the class of states given by eq.\ \ref{nonlocal}, $\omega$ can be expressed explicitly as a function of $\theta$, since one can easily calculate $\mathcal{C}(\ket{\psi_{\theta}}\bra{\psi_{\theta}}) = \sin(2\theta)$, thus,
\begin{equation}
\label{probthe}
   \omega(\theta) = \frac{1}{2}+\frac{1}{4} \sqrt{1 + \sin^2(2 \theta)}.
\end{equation}

It can be easily checked that both the maximum values for both the winning probability and concurrence are achieved for  $\theta={\pi}/{4}$, i.e.\ $\omega{}(\pi/4)\approx 0.85...$ and  $\mathcal{C}(\pi/4)=1$, which (as expected) corresponds to a maximally entangled state, namely $\ket{\phi^+} = \ket{\psi_{\pi/4}}$.

\subsection{Device-independence and self-testing}
The concept of self-testing, first introduced by Mayers and Yao \cite{MY98}, allows one to consider a device-independent scenario, where an adversary is given complete control of the manufacture of the devices, on top of the usual security assumptions against malicious quantum agents.
For instance the devices may have memories, clocks, location tracking, or any other form of hidden mechanism \cite{AMPS16}.

We use the same assumptions for device-independence as in \cite{AMPS16}.
In principle, the interactions between the parties could be modeled as black boxes with classical inputs and outputs, where entering an input must result in an output.
Each party has access to a trusted and private source of randomness inside of their laboratory, which is assumed to be perfectly isolated (no leaks to the environment).
Furthermore, the parties are restricted by the laws of quantum mechanics and the communication between them may be prevented at will.
(Note: this is a weaker condition than the usual non-signaling in a fully fledged Bell scenario \cite{PABGMS09}.)

We adopt the terminology of \cite{SB20}, where in a bipartite device-independent scenario one aims to test or certify an unknown \textit{physical experiment} (\(\{\rho_{AB},\Pi_A, \Pi_B\}\)) consisting of a \textit{physical state} (\(\rho\)) (inside a black-box) and \textit{physical measurements} (settings of the black-box),  by comparing the statistics outputted by the black-box against the expected statistics predicted by some predefined \textit{reference experiment}  \(\{\ket{\psi}_{AB}, {\Pi'}_A, {\Pi'}_B\}\).

Under the assumptions of device-independence, we write the expression for the probabilities of the physical experiment as follows, 
\begin{equation}
\label{eq:di-generic}
    \Pr(a,b|s,t)_{\textup{Box}} = \text{Tr}[\rho_{AB} (\Pi_A^s \otimes \Pi_B^t)].
\end{equation}

If we consider the purification of the physical state \(\rho\) (\cite{SB20}), we can still write it as
\begin{equation}
    \Pr(a,b|s,t)_{\textup{Box}} = \bra{\psi}(\Pi^s_{A}\otimes \Pi^t_{B}\otimes \mathbb{1}_E )\ket{\psi}_{ABE}.
\end{equation}
Then, according to the usual notion of self-testing, if the previous correlations \(\Pr(a,b|s,t)_{\textup{Box}}\) allow Alice and Bob to infer the existence of a reference experiment \(\{\ket{\psi}_{AB}, {\Pi'}_A, {\Pi'}_B\}\)  uniquely achieving such correlations --- up to local isometries and ancillary degrees of freedom --- we say that Alice and Bob successfully self-tested the physical experiment against reference experiment.\\

\paragraph*{Device-independence by self-testing via simulation \cite{K16, SB20}:}
An alternative approach to proving device-independence by self-testing is via simulation.
It states that the physical experiment being tested should only need to simulate some desired statistical behavior of the reference experiment (i.e.\ the protocol). This interpretation is intimately related to extractability using quantum channels (see \cite{CKJS19}), which is  proven to be equivalent to the previously mentioned and more familiar notion of self-testing using local isometries and tracing out the extra degrees of freedom \cite[Proof Appendix. A]{CKJS19}.

This notion of self-testing via simulation is naturally applied to our case.
For our purposes, it will suffice that the CHSH self-test certifies the capability that the black-box can simulate a certain amount of entanglement correlations specified by the reference experiment.
This comes in opposition to certifying the physical contents within the black-box against the reference experiment.

Therefore, although we make no assumption on the physical experiment (\(\{\ket{\psi}_{ABE}, \Pi^s_{A},  \Pi^t_{B}\}\)), we assume (without loss of generality) the following constraints on the reference experiment: 
only \(2\otimes 2\) dimensional pure reference states and projective measurements are considered, i.e.\ \(\ket{\psi}_{AB}\) becomes a pure partially entangled two-qubit state \(\ket{\psi_{\theta}}_{AB}\), as defined in eq.\ \ref{nonlocal} (up to a local change of basis), and \(\Pi'\) becomes a rank 2 projector, \(P\).
We are motivated to apply these restrictions since we know, from Cleve et al. \cite{CHTW04}, that having a state with higher dimensional entanglement than that of a Bell pair gives no advantage in bi-partite binary non-local games of sufficiently small dimensions (like the CHSH game), and also that projective measurements (\(P\)) are sufficient to achieve the Tsirelson bound \cite{T80} in such cases.

Thus, under these constraints and according to the notion of self-testing via simulation using the CHSH game, we can write, 
\begin{equation}
\label{sim}
    \Pr(a,b|s,t)_{\textup{Box}}\cong \bra{\psi_{\theta}}(P^s_{A}\otimes P^t_{B})\ket{\psi_{\theta}}_{AB}.
\end{equation}
That is, if the statistics provided by the black-box (LHS of eq.\ \ref{sim}) simulate the correlations achieved by implementing the best strategies of the reference experiment  (RHS of eq. \ref{sim}), then we certify a lower bound on the entanglement correlations that the physical experiment (i.e the black-box) can simulate.
Thus, in this case, testing the physical experiment through CHSH tests does not certify the physical state inside the box, but rather a lower bound on the entanglement correlations that the physical state can simulate (up to 1 \textit{ebit})\footnote{If we intended on certifying the physical state itself, that would require the use of the so-called tilted Bell inequalities \cite{YN13,AMPS12}.}.
Notice that, under the constraints of the device-independent setting, we are excluding the possibility that such correlations are simulated by local hidden variable models, and as such, these correlations are intrinsically non-local.
As we will see, this is enough for the purposes of the quantum authorization protocol  we consider.
In particular, when assuming that the reference measurements are optimal for a given value of the concurrence, the statistics used to verify that the physical experiment simulates the reference experiment (i.e.\ the RHS of eq.\ \ref{sim}) are given by eq.\ \ref{prob} (or eq.\ \ref{probthe} explicitly depending on $\theta$).

\section{Quantum authorization}
We now introduce our quantum authorization protocol.
Briefly, a first party, the \autho{}, must be able to uniquely establish authorization levels for another party, the \users{}, by playing a series of CHSH games with each of them.
Such \users{} will be grouped into discrete hierarchical levels, and granted authorization to privileged information depending on their level.
Then, \users{} assigned to a specific level must not be able to cheat and pretend they are in a higher one in order to access unprivileged information.
The levels assigned to the \users{ } are defined by a mapping of the \autho{}'s choosing: assign a specific amount of entanglement to some authorization level, such that higher amounts of entanglement must always correspond to higher levels of authorization.
Therefore, the authorization level is encoded in the amount of entanglement present in the quantum resources distributed by a \distrib{} to the \users{}.
Afterward, the resources are used by both the \autho{} and \user{} when playing a series of CHSH games to quantify the amount entanglement present, and authorization is granted.

Let \(\mathtt{DB}\) be a totally ordered database which is divided from \(1\) to \(\ell\) in increasing levels of access permission \(L_i\), as \(\mathtt{DB} = \{L_1, L_2, \dots, L_{\ell}\}\).
So, a \user{} which proves that it has right to the level \(k\), should only be able to access the segment of the database corresponding to \(\mathtt{DB}_k = \bigcup_{i=1}^k L_i\).

We define an authorization algorithm \(\mathsf{A}\) such that \(\mathtt{DB}_k \leftarrow \mathsf{A}(\mathcal{C}_i)\), meaning that given a unique concurrence value \(\mathcal{C}_i\), it returns the corresponding segment of the database up to level \(k\) (matching that \(\mathcal{C}_i\)).
For this, the database is split according to prespecified concurrence levels \(\{\mathcal{C}_1,\mathcal{C}_2,\dots,\mathcal{C}_{{\ell}}\}\) defined by the \autho{} respecting the total order of the set \(\mathtt{DB}\), that is, $\mathcal{C}_i > \mathcal{C}_j$ iff $L_i > L_j$.
Now, we have $\theta_i = \arcsin{(\mathcal{C}_i})/2$ as the angle for which $\ket{\psi_{\theta_i}}$ has concurrence $\mathcal{C}_i$, and $\omega_i= (1+\sqrt{1+{\mathcal{C}_i}^2})/2$ the optimal winning probability to win the CHSH when using non-local resource $\ket{\psi_{\theta_i}}$.

Following, we present a thorough description of the protocol.
In a setup phase, the \distrib{}, distributes entangled pairs of qubits between each \user{} and the \autho{}.
The pairs given to the \user{} should be created according to its level of authorization established beforehand.
Note that the \users{} not only have knowledge of their own respective level $k$, but also know the correspondence between the amounts of entanglement for each level in the hierarchy, which might be public.
Moreover, it is possible for the role of \distrib{} and \autho{} to be played by the same entity, but throughout the presentation we will assume they are different parties for generality purposes.
Finally, we do not consider quantum noise effects in the channels shared between the \users{} and the \autho{}, as such we do not need to consider robust forms of self-testing of the CHSH game/inequality.

Schematically, the protocol goes as follows, for prespecified \(\niter\) iterations:
\begin{enumerate}
    \item The \user{} starts the interaction by sending its identifier to the \autho{} to request authorization to some level \(k\).
    \item The \user{} and the \autho{} each sample the question for the upcoming CHSH game.
    \item \label{en:protocolreply} The \user{} and \autho{} play the CHSH game by measuring its qubit following the best strategy for level $k$, maximizing their winning probability (eq.\ \ref{prob}).
    \item The \user{} and the \autho{} announce their inputs and outputs. Then, check their validity regarding the winning condition, \(s\cdot t = a\oplus b\).
    \item Steps 2-4 are repeated $N$ times, and the \autho{} compares the number of games won in relation to the expected number for the level \(k\). If there is no match (up to some predefined error \(\varepsilon\)), it aborts. Otherwise, it provides authorization to the computed level, \(\mathtt{DB}_k\).
\end{enumerate}

The protocol is parameterized by the level discrimination error, \(\varepsilon\), and, the levels must be set up such that the distributions of any two levels do not overlap in agreement with \(\varepsilon\). 
This error can be made arbitrarily small, given that there are enough entangled qubits between the \autho{} and the \user{} to iterate the CHSH game.
In fact, all pairs of qubits shared between the \autho{} and the \user{} should have the same level of entanglement and be independent, i.e.\ separable.
Thus, the CHSH games can be seen as a random variable following a binomial distribution, for \(\niter\) played games (i.e.\ \(\niter\) Bernoulli random variables), with winning probability \(\omega\).
Mind that, by comparing the number of games won, the \autho{} is analysing the entanglement level that was present in the qubits it shared with the \user{}.
This comes from the fact that the maximum winning probability and the concurrence are related by a bijective function (eq.\ \ref{prob}), so we will use these two descriptions interchangeably.

\paragraph*{Device-independence:}
Our quantum authorization protocol holds on the fact that the \user{} is able to show that it holds certain non-local resources (given to it beforehand), by playing multiple CHSH games and estimating the concurrence level of the resources used while playing the games.
Indeed, in step \ref{en:protocolreply} of the protocol, the CHSH games are played by sampling the questions and setting the answers.
These are announced at a prespecified instance in time, which is key regarding the device independence of the protocol.
Surely, the parties do not need to assume what they have in the boxes:
If the boxes do not have inside a physical state capable of simulating the reference experiment, i.e.\ achieving a certain number of games won, then the CHSH bound given the optimal strategies for that concurrence level (eq.\ \ref{prob}) will not be reached and the protocol will abort, implying that the physical experiment fails to simulate the reference experiment.
In fact, as previously mentioned, we are doing a device-independence certification of a lower-bound on the entanglement of the physical state, which serves the purpose of our authorization protocol.
This is similar to the usual verification step of the non-locality done as a subroutine in device-independent QKD protocols \cite{VV14}.

\paragraph*{Proof of security:}
We will use the tail inequalities for the Binomial distribution, e.g. Chernoff bound \cite{C52}, to guarantee that the probability of the random variable deviating from the mean decreases exponentially with the number of Bernoulli experiments, \(\niter\).

\begin{lemma}[Chernoff bound \cite{C52, AB09}] Let \(X_1, X_2, \dots, X_\niter\) be independent Bernoulli random variables, then,
    \begin{equation*}
        \Pr[\left|\sum_{i=1}^\niter X_i - \mu\right| \geq c\mu] \leq 2^{-c^2 \niter/2}
    \end{equation*}
    for every \(c>0\), where \(\mu = \sum_{i=1}^{\niter}\mathbb{E}(X_i)\). 
    \label{lemma:chernoff}
\end{lemma}

\begin{theorem}
    The quantum authorization protocol is unconditionally secure for the \autho{}.% against a malicious \user{}.
\end{theorem}

\begin{proof}(\textit{Soundness})
    First, before announcing the outcomes of the CHSH games, the \user{} cannot know which games it won or lost.
    Since this announcement is done simultaneously by the parties, the \user{} is not able to wait for the \autho{} to publish its result and then make up its own.
    So, a \user{} must play the game according to some (possibly malicious) strategy and announce its inputs and outputs, or the \autho{} will abort.
    Finally, it has no way of knowing which answers lead to a win (the distribution depends on the chosen strategy), so the probability of choosing a set of winning results decreases exponentially with the size of this set (i.e. with \(N\)).
    
    Second, from Lemma \ref{lemma:chernoff}, \(\varepsilon\) can be adequately chosen in order to make the interval \((\mu - \varepsilon, \mu+\varepsilon)\) small enough, such that a \user{} which cannot win with probability \(\omega\) will fall outside this interval, with overwhelming probability, exponential in \(N\).
    
    Lastly, if a \user{} intended for level \(k\) requests access to $k'>k$, the \autho{} will implement an optimal strategy for $k'$.
    Nonetheless, this is inconsequential as the entanglement shared is not enough to realize it, i.e. given the total order previously established for \(\{\mathcal{C}_1,\mathcal{C}_2,\dots,\mathcal{C}_{{\ell}}\}\) (\(\mathcal{C}_{i}\) corresponding to level \(i\)), and regarding the optimal winning probability (eq. \ref{prob}), we have \(\omega(\mathcal{C}_{k'}) > \omega(\mathcal{C}_{k}) \text{ iff } k' > k\) \cite{V02}.
    Again from Lemma \ref{lemma:chernoff}, for \(N\) iterations, the probability of falling in the interval \((\mu - \varepsilon, \mu+\varepsilon)\) is exponentially small.
    
    \textit{Impossibility of collusion:} No set of \users{} can join together and impersonate another \user{}.
    This is a consequence of the aforementioned impossibility of entanglement distillation without collaboration from the \autho{}.
    By forfeiting the possibility of LOCC between the \users{} and the \autho{}, collaboration between the \users{} provides them no advantage in breaking security.
\end{proof}

\begin{theorem}
    The quantum authorization protocol is unconditionally secure for the \user{}.%against a malicious \autho{}.
\end{theorem}

\begin{proof}(\textit{Completeness})
    This comes as a trivial consequence of the way the protocol is set.
    Indeed, the \user{} taking part in the protocol execution knows exactly which is the level (\(k\)) that it should have access to.
    So, if the \user{} acts honestly and the \autho{} aborts the execution or grants it access to some other level \(k' \neq k\), the \user{} aborts the execution as it knows \autho{} is attempting to cheat.
    Indeed, security here is unconditional, as this is the same as assuming the \autho{} is a trusted party: either the \user{} is granted access to its preestablished level, or it aborts.
    
    Note that this trivial security is a consequence from the heavily asymmetric nature of the protocol, where the \autho{} has basically all the power (it owns \(\mathtt{DB}\)), and a \user{} is challenged to access it.
    However, to protect the \users{}, once granted, the \autho{} cannot revoke their access level without being caught.
\end{proof}

\paragraph*{Implementation and parameter instantiation:}
We implement a proof-of-concept of the protocol in Python using the Qiskit open-source framework \cite{qiskit}, which we provide as open-source in \href{https://github.com/goulov/qauth}{https://github.com/goulov/qauth}.
In order to instantiate the parameters for a practical implementation of the protocol, one must explicitly state the total number of levels \(\ell\), and for each level assign a concurrence value (splitting the interval evenly) implying a different probability of winning each game.
Given this winning probability, one can use the Chernoff bound (Lemma \ref{lemma:chernoff}), for a security parameter \(\lambda\) and suitable \(c\) (deviation from the expected number of wins) to obtain the required number of iterations \(N\) for that security parameter.
Indeed, given \(\ell\), then \(c = 1/(2\ell)\) and \(N=(2\lambda)/(c^2)\) and \(\varepsilon = c\mu\).
We provide concrete values for the usual \(128\)-bit and \(256\)-bit security for different number of access levels \(\ell\) (Table \ref{tab:parameter}).\\

\begin{table}[h]
    \centering
    %\resizebox{.6\linewidth}{!}{
    \caption{Instantiation of the parameters for security \(2^{-128}\), and \(2^{-256}\).}
    \begin{tabular*}{\linewidth}{@{\extracolsep{\fill}} c c c c c}
        \hline
        \(\ell\) & \(N_{128}\) & \(\varepsilon_{128}\)& \(N_{256}\) &  \(\varepsilon_{256}\)\\
        \hline
        2 & 4096 & 768 & 8192 & 1536\\
        4 & 16384 & 1536 & 32768 & 3072\\
        6 & 36864 & 2304 & 73728 & 4608\\
        \hline
    \end{tabular*}
    %}
    \label{tab:parameter}
\end{table}

\section{Conclusions}
In this work we presented a simple quantum authorization primitive based on non-locality.
The protocol grants a party, the \autho{}, the power to attribute authorization levels from a discrete hierarchy, to a number of other parties, the  \users{}. 
Thus, each individual \user{} will have access to a subset of a private database in accordance to its respective authorization level.
The authorization level of each \user{} is encoded (non-locally) in the amount of entangled information shared with the \autho{}, formalized through the entanglement concurrence, which is later verified by the fraction of CHSH games the \users{} and \autho{} win.
Several properties pertaining to the  security of the protocol are exhibited and proved, namely, that the protocol is sound and complete. Furthermore, we cast the protocol within a device-independent setting, where it is shown that the protocol can be entirely reasoned as if being implemented with black-boxes, wherein the behaviour of such boxes is subsequently self-tested  by the CHSH game/inequality via simulation. In particular, we certify a lower-bound on the entanglement up to 1-\textit{ebit}, which suffices for the purposes of proving device-independence in our protocol. 
Finally, we instantiated the parameters for the usual security levels and implemented the protocol, using the Qiskit open-source framework, which we made publicly available online.

\section*{Acknowledgements}
The authors thank the support from DP-PMI and FCT (Portugal). RF and MG acknowledge grants PD/BD/128636/2017 and PD/BD/135182/2017, respectively.

Both authors contributed equally.

\bibliographystyle{ieeetr}
\bibliography{bib}

\end{document}